\documentclass[%
twocolumn,%
secnumarabic,%
amsmath,%
amssymb,%
nobibnotes,%
aps,%
prl,%
superscriptaddress%
]{revtex4-1}

\newcommand{\str}[1]{{#1}^\ast}
\newcommand{\prm}[1]{{#1}^\prime}
\newcommand{\pprm}[1]{{#1}^{\prime\prime}}

\newcommand{\meffs}{m_\mathrm{eff}^2}

\newcommand{\bk}{\mathbf{k}}
\newcommand{\bx}{\mathbf{x}}

\newcommand{\ket}[1]{\left|#1\right\rangle}

\newcommand{\bracket}[3]{\left\langle#1\middle|#2\middle|#3\right\rangle}

\usepackage{lipsum}
\usepackage{graphicx}

\begin{document}

\title{Novel quantum initial conditions for inflation}%

\author{W.J.\ Handley}%
\email{wh260@mrao.cam.ac.uk}
\affiliation{Astrophysics Group, Cavendish Laboratory, J.~J.~Thomson Avenue, Cambridge, CB3 0HE, UK }
\affiliation{Kavli Institute for Cosmology, Madingley Road, Cambridge, CB3 0HA, UK }
\author{A.N.\ Lasenby}%
\email{a.n.lasenby@mrao.cam.ac.uk}
\affiliation{Astrophysics Group, Cavendish Laboratory, J.~J.~Thomson Avenue, Cambridge, CB3 0HE, UK }
\affiliation{Kavli Institute for Cosmology, Madingley Road, Cambridge, CB3 0HA, UK }
\author{M.P.\ Hobson}%
\email{mph@mrao.cam.ac.uk}
\affiliation{Astrophysics Group, Cavendish Laboratory, J.~J.~Thomson Avenue, Cambridge, CB3 0HE, UK }

\begin{abstract}
  We present a novel approach for setting initial conditions on the mode functions of the Mukhanov-Sazaki equation. These conditions are motivated by minimisation of the renormalised stress-energy tensor, and are valid for setting a vacuum state even in a context where the spacetime is changing rapidly. Moreover, these alternative conditions are potentially observationally distinguishable. We apply this to the kinetically dominated universe, and compare with the more traditional approach. 
\end{abstract}

\date{\today \hspace{20pt} Accepted by Physical Review D}%
\maketitle

\section{Introduction}
Traditionally, quantum initial conditions for inflation are set using the Bunch-Davies vacuum. This approach is valid in de-Sitter space and other asymptotically static spacetimes. Rapidly evolving spacetimes, however, do not admit such an easy quantisation. 

In a recent work~\cite{Handley+2014}, we showed that the classical equations of motion suggest that the universe in fact emerged a rapidly evolving state, with the kinetic energy of the inflaton dominating the potential in a pre-inflationary phase. 
This can be used to set initial conditions on the background variables such as the inflaton value and Hubble parameter. 
In order to make contact with real observations, the effect that this phase has on the primordial power spectrum requires a semi-classical quantum mechanical treatment of the comoving curvature perturbation.

Hamiltonian diagonalisation is the simplest approach for setting quantum initial conditions in a general spacetime, and derives the vacuum from the minimisation of the Hamiltonian density. This approach has been criticised in the past as it does not admit a consistent interpretation in terms of particles~\cite{Fulling+1989,Fulling_HD}. Other approaches such as the adiabatic vacuum go some way to rescuing the particle concept, but have additional theoretical issues. 

The issue of the particle interpretation stems from an attempt to apply a Minkowski spacetime concept outside the region of its validity.  
We postulate that the minimisation of an energy density is still an appropriate way to define a vacuum. In order to avoid the issues raised against Hamiltonian diagonalisation, we motivate our initial conditions from the minimisation of the {\em renormalised\/} stress-energy density. 
Indeed, if one takes care to minimise the correct quantity (using the theory of quantum fields in curved spacetime), then novel initial conditions can be derived which differ from the traditional Hamiltonian diagonalisation conditions.

After the relevant background material is reviewed, we develop a generic mechanism for setting initial conditions. These reduce to the Bunch-Davies case in asymptotically static spacetimes (such as de-Sitter space), but yield different results otherwise. The aim is that these should be more theoretically robust. Additionally, these conditions are potentially distinguishable using observational data.

We then apply this procedure to the kinetically dominated universe, but delay the observational analysis to a later work.

\section{Background}

We denote a general action via
\begin{equation}
  S_I = \int d^4x\sqrt{|g|}\mathcal{L}_I,
  \label{eqn:general_action}
\end{equation}
where $\mathcal{L}_I$ is the {\em Lagrangian density}. We work in natural units $\hbar=c=1$ and set the reduced Planck mass $m_\mathrm{p} = {(8\pi G)}^{-1/2} = 1$. Dots denote differentiation with respect to cosmic time $\dot{f}\equiv \frac{d}{dt}f$, and primes denote differentiation with respect to conformal time $\prm{f}\equiv\frac{d}{d\eta}f$.

We begin by briefly summarising the classical theory of cosmological perturbations for a general scalar field, before discussing the quantisation of such a theory

\subsection{The classical action}
Consider~\cite{Baumann+2009} a canonical scalar field $\phi$ minimally coupled to gravity $S= S_G + S_\phi$ with:
\begin{equation}
  \mathcal{L}_G = \frac{1}{2}R, 
  \qquad
  \mathcal{L}_\phi = \frac{1}{2}g^{\mu\nu}\nabla_\mu\phi\nabla_\nu\phi - V(\phi),
  \label{eqn:action}
\end{equation}
Extremising this action with respect to the fields $\phi$ and $g_{\mu\nu}$ recovers the Klein-Gordan and Einstein equations respectively:
\begin{align}
  \left( g^{\mu\nu}\nabla_\mu\nabla_\nu + \frac{dV}{d\phi} \right) \phi &= 0,
  \label{eqn:klein_gordon}\\
  G_{\mu\nu}\equiv R_{\mu\nu}-\frac{1}{2}g_{\mu\nu}R&= T_{\mu\nu},
  \label{eqn:einstein}
\end{align}
where the stress-energy tensor is:
\begin{equation}
  T_{\mu\nu} = \nabla_\mu\phi \nabla_\nu\phi - \frac{1}{2}g_{\mu\nu} \nabla_\alpha\phi \nabla^\alpha\phi +g_{\mu\nu} V(\phi).
  \label{eqn:SET}
\end{equation}

In cosmology, we assume that at zeroth order both the metric $g_{\mu\nu}$ and scalar field $\phi$ are homogeneous and isotropic. Applying these assumptions to equations~(\ref{eqn:klein_gordon})~\&~(\ref{eqn:einstein}), we find:
\begin{align}
  \dot{H}+H^2 &= -\frac{1}{3}\left( \dot{\phi}^2 - V(\phi) \right),
  \label{eqn:Raychaudhuri}\\
  0&=\ddot{\phi} + 3H\dot{\phi} + \frac{dV}{d\phi},
\end{align}
where the Hubble parameter $H = \dot{a}/a$.

\subsection{Inflationary perturbations}
One then considers perturbations about these background solutions 
\begin{align}
  \phi=&\phi(t) + \delta\phi(t,x),
  \label{eqn:phi_perturbation} \\
  ds^2 =& {(1+2\Phi)}dt^2 + 2a\left( \partial_i B-S_i \right)dx_i dt \nonumber\\
  &+a^2{(1-2\Psi)}\delta_{ij}dx_i dx_j \nonumber\\
  &+\left( 2\partial_i\partial_j E + 2\partial_{(i\phantom)}F_{\phantom(j)} + h_{ij} \right)dx_i dx_j,
\end{align}
where without loss of generality, the vector fields $S_i,F_i$ are divergenceless, and the tensor field $h_{ij}$ is symmetric, divergenceless and traceless.

We are interested in the gauge-invariant co-moving curvature perturbation:
\begin{equation}
  \mathcal{R}\equiv \Psi - \frac{H}{\dot{\phi}}\delta\phi,
\end{equation}
since it is this quantity which defines the primordial power spectrum for seeding cosmological perturbations. Working in the co-moving gauge $\delta\phi=0$, and expanding the action $S$ to second order in $\mathcal{R}$, gives:
\begin{equation}
  S_{(2)} =  \int d^4 x a^3\frac{\dot{\phi}^2}{H^2}{\left[ \dot{\mathcal{R}}^2 - a^{-2} {\left( \partial_i\mathcal{R} \right)}^2 \right]}.
\end{equation}
Note that the dependence on $V(\phi)$ is implicit in the variables $H,\dot{\phi},a$ and $\mathcal{R}$.
Defining the Mukhanov variable,
\begin{equation}
  v = z\mathcal{R},\qquad z=\frac{a\dot{\phi}}{H},
  \label{eqn:mukhanov_variable}
\end{equation}
and transforming $t$ into conformal time $\eta = \int^t d\tau/a(\tau) $ yields:
\begin{equation}                                 
  S_{(2)} =  \int d\eta d^3 x {\left[ {\left( \prm{v} \right)}^2 - {\left( \partial_i v \right)}^2 + \frac{\pprm{z}}{z} v^2 \right]}.
  \label{eqn:v_action}
\end{equation}
This is the canonically normalised action for a scalar field with time-dependent ``effective'' mass $\meffs = -\pprm{z}/z$.

\section{Quantisation via Hamiltonian Diagonalisation}                                
We now consider the traditional quantisation of the action~(\ref{eqn:v_action}) via Hamiltonian diagonalisation. This is a standard method in the inflationary literature, but has several theoretical issues which will be discussed. To begin, one writes;
\begin{equation}
  v = \int \frac{d^3 k}{{(2\pi)}^3} \left[ a_\bk \chi_\bk(\eta)e^{i\bk\cdot\bx} + a_{\bk}^{\dagger} \str{\chi_\bk}(\eta)e^{-i\bk\cdot\bx} \right], 
  \label{eqn:v_quant}
\end{equation}
which expresses the operator $v$ as a superposition of creation and annihilation operators $\{a_\bk,a_\bk^\dagger\}$~\cite{Mukhanov+2007}, with the mode functions written in separated form ${u_\bk = \chi_\bk(\eta) e^{i\bk\cdot\bx}}$. If one requires that the scalar field satisfies the equations of motion, and that canonical commutator relation:
\begin{equation}
  [ a_\bk^{\phantom\dagger} a_{\prm{\bk}}^{\dagger} ] = {(2\pi)}^3\delta^{(3)}(\bk-\prm{\bk}),
  \label{eqn:commutator}
\end{equation}
holds true, then the temporal part $\chi_\bk(\eta)$ of the mode functions $u_\bk$ must satisfy:
\begin{align}
  \pprm{\chi_\bk} + \left( k^2 - \frac{\pprm{z}}{z} \right)\chi_\bk &= 0,
  \label{eqn:mode_mukhanov}
  \\
  \prm{\chi_\bk} \str{\chi_\bk} - \prm{\str{\chi_\bk}} \chi_\bk &= -i.
  \label{eqn:normalisation}
\end{align}
The first of these is the classical equation of motion of the action~(\ref{eqn:v_action}), whilst the second is a normalisation constraint.

\subsection{Choosing a vacuum}
The complex mode functions $\chi_\bk$ are not fully determined by condition~(\ref{eqn:normalisation}). Although the overall phase of the mode $\chi_\bk$ is unimportant, there is an additional degree of freedom for each $\bk$ to be determined. The choice of this is equivalent to choosing a vacuum state $\ket{0}$, defined by $a_\bk\ket{0}=0$.

The traditional approach is to consider the Hamiltonian of the Mukhanov variable, which after normal ordering takes the form:
\begin{align}
  H = \frac{1}{2}\int\frac{d^3k}{{(2\pi)}^3} 
  &\Big[ a_\bk a_{-\bk} F_\bk(\eta) + a_\bk^\dagger a_{-\bk}^\dagger \str{F_\bk}(\eta) \nonumber \\
  &+ \left(2a_\bk^\dagger a_\bk + \delta^{(3)}(0)\right)E_\bk(\eta) \Big], 
\end{align}
where
\begin{align}
  E_\bk(\eta) &= |\chi_\bk^\prime|^2 + \omega_k^2|\chi_\bk|^2,\quad
  F_\bk(\eta) = {\chi_\bk^\prime}^2 + \omega_k^2{\chi_\bk}^2,\\
  \omega_k^2(\eta) &= k^2-\frac{\pprm{z}}{z}.
  \label{eqn:def_omega}
\end{align}
It is therefore attractive to choose either (i) the vacuum as an eigenstate of the Hamiltonian:
\begin{equation}
  H\ket{0}\propto\ket{0} \Rightarrow F_\bk=0,
  \label{eqn:eigenstate}
\end{equation}
or (ii) that the vacuum minimises the expected energy:
\begin{equation}
  \bracket{0}{H}{0} \propto \int \frac{d^3k}{{(2\pi)}^3} E_\bk.
  \label{eqn:hamil}
\end{equation}
As can be shown with standard linear algebra, these two conditions are equivalent, and result in the requirement that:
\begin{equation}
  |\chi_\bk|^2 = \frac{1}{2\omega_k}, \qquad \prm{\chi_\bk} = -i \omega_k \chi_\bk.
  \label{eqn:hd_condition}
\end{equation}
which provides enough information to set unambiguous initial conditions for the mode equation~(\ref{eqn:mode_mukhanov}).
When the condition~(\ref{eqn:hd_condition}) is satisfied, the Hamiltonian is diagonalised such that:
\begin{equation}
  H(\eta_0) = \int\frac{d^3k}{{(2\pi)}^3} 
  \left(a_\bk^\dagger a_\bk + \frac{1}{2}\delta^{(3)}(0)\right)\omega_k(\eta_0). 
  \label{eqn:diag_hamil}
\end{equation}
We henceforth refer to~(\ref{eqn:hd_condition}) as the Hamiltonian diagonalising (HD) vacuum choice.

From~(\ref{eqn:diag_hamil}), one may easily show that $a_\bk^\dagger\ket{0}$ is a state with energy $\omega_k(\eta_0)$ and momentum $\bk$. One therefore traditionally interprets the action of $a_\bk^\dagger$ at time $\eta_0$ as creating a ``particle'' from the vacuum. This is a well established interpretation in flat Minkowski space. %In curved spacetime however, we shall see in the following section that this interpretation becomes troublesome.

Note that in general, the conditions~(\ref{eqn:hd_condition}) are only satisfied at a specific time $\eta_0$, and the vacuum is thus a time-dependent notion. 
Indeed, it is straightforward to show that at some different time $\eta_1$, the expectation $\bracket{0}{H(\eta_1)}{0}$ of the Hamiltonian in the ground state $\ket{0}$ is in general larger than the minimum possible value at $\eta_1$. This is interpreted as the vacuum state $\ket{0}$ at $\eta_0$ containing particles at other times $\eta_1$.

Thus, if one sets these conditions at a time $\eta_0$, one is effectively setting the universe to be in the vacuum state at that time. The expansion of the universe then excites the vacuum, creating ``particles'' at later times $\eta_1$. The question as to what is the ``correct'' $\eta_0$ at which to set these conditions is currently an unresolved theoretical (or indeed observational) issue.

\subsection{Criticism of Hamiltonian diagonalisation}

Astute readers will have spotted that the expected energy~(\ref{eqn:hamil}) is divergent. Whilst the implicit $\delta^{(3)}(0)$ in the proportionality constant of~(\ref{eqn:hamil}) is harmless, and merely accounts for the contribution from the infinite volume of space, there is a second divergence which requires closer attention. For large $k$, $E_\bk\sim \omega_k \sim k$, and hence the integral~(\ref{eqn:hamil}), which represents the energy density, is ultraviolet divergent as $k^4$. 

In traditional quantum field theory, this divergence is subtracted as one only measures energy differences. This is also applicable to spacetimes that are asymptotically static (such as de-Sitter space).
However, in changing spacetimes, where the vacuum is time dependent, this subtraction can only be performed at a single instant. If one then advances in time by some finite amount, the space-time generates an infinite particle density~\cite{Fulling_HD,Fulling+1989}. 

This is clearly unphysical, causing some authors~\cite{Fulling_HD} to discard Hamiltonian diagonalisation as an inappropriate methodology for choosing a vacuum state.

\section{Alternative Quantisations}
The particle concept can be somewhat rescued by considering the adiabatic vacuum. This is well defined when the spacetime is changing slowly, as one can then perform an adiabatic expansion. The $n$\textsuperscript{th} order adiabatic vacuum at time $\eta_0$ is defined by matching the general solution onto the $n$\textsuperscript{th} order adiabatic expansion at time $\eta_0$. This has the satisfying property of more closely corresponding to what a freely falling particle detector would measure, and is generally agreed to be superior to Hamiltonian diagonalisation.

However, there are still some issues with this vacuum. First, it is only usable in slowly changing spacetimes, so only goes halfway to solving the general problem. Second, it introduces a further ambiguity in vacuum choice, namely that of which value of $n$ to choose. Since the adiabatic expansion is asymptotic, it does not in general converge for large $n$. One must pick a specific term of the series to truncate at, and there is little theoretical guidance as to what value $n$ to choose.

We believe that the adiabatic vacuum is in fact trying to rescue the particle concept unnecessarily. A particle interpretation is doomed to failure in general curved spacetime because of the global nature of their definition. Particles are defined in terms of field modes over a large patch of the manifold. Whilst for higher $\bk$ modes the environment looks effectively Minkowksi, low $\bk$ modes are sensitive to the large scale structure of spacetime.

It would be more sensible to base the notion of a vacuum not in terms of a ``particle-less'' state, but in terms of the minimisation of a {\em local energy density}, such as the $0$-$0$ component of the stress-energy tensor. Unfortunately being quadratic in the field $\phi$, like the Hamiltonian, $\bracket{0}{T_{00}}{0}$ is also divergent.

In order to ameliorate this difficulty, we must adopt a more sophisticated approach. 

\section{Quantum fields in curved spacetime} 
This is the semi-rigorous theory of fields in which gravity is strong enough to generate curvature, but the quantum mechanics only affects spacetime to low order. It can therefore be thought of as a one-loop approximation to quantum gravity.

Traditionally~\cite{Birrell+1984,Parker+2009}, one considers a scalar field Lagrangian with mass $m$, with action:
\begin{equation}
  S = \int d^4x \sqrt{|g|}\left( \frac{1}{2}g^{\mu\nu}\nabla_\mu\phi\nabla_\nu\phi - \frac{1}{2}m^2\phi^2 \right),
  \label{eqn:scalar_field}
\end{equation}
where for simplicity we are considering the case of minimal coupling $\xi=0$.  
In the context of FRW spacetime, the modes are quantised as:
\begin{equation}
  \phi(x) = \int\frac{d^3 k}{{(2\pi)}^3a(\eta)} \left[ a_\bk \chi_\bk(\eta)e^{i\bk\cdot\bx} + a_{\bk}^{\dagger} \str{\chi_\bk}(\eta)e^{-i\bk\cdot\bx} \right],
  \label{eqn:mode_expansion}
\end{equation}
where the mode functions are written in separated form $u_\bk = a{(\eta)}^{-1}\chi_\bk(\eta)e^{i \bk\cdot\bx}$. The additional conformal factor of $a{(\eta)}^{-1}$ generates mode equations without first order derivatives in $\eta$.
Requiring that the scalar field satisfies the equations of motion, and that the commutation relation~(\ref{eqn:commutator}) remain true,
one finds that the mode functions $\chi_\bk$ must satisfy:
\begin{align}
  \pprm{\chi}_\bk + \left[ k^2 + a^2 m^2 - \frac{\pprm{a}}{a}  \right] \chi_\bk &= 0,
  \label{eqn:chi_equation}\\
  \prm{\chi_\bk} \str{\chi_\bk} - \prm{\str{\chi_\bk}} \chi_\bk &= -i
  \label{eqn:normalisation_1}
\end{align}

\subsection{Application to inflation}
The similarity between equations~(\ref{eqn:mode_mukhanov}) and~(\ref{eqn:chi_equation}) is striking. It suggests solving for the quantum curvature perturbation is equivalent to solving a massless scalar field in an alternative spacetime with scale factor satisfying:
\begin{equation}
  \frac{\pprm{a}}{a} = \frac{\pprm{z}}{z}.
  \label{eqn:mapping}
\end{equation}
This may be explicitly solved for $a(\eta)$ as:
\begin{equation}
  a(\eta) = A\:z(\eta) + B\:z(\eta) \int^\eta \frac{dx}{{z(x)}^2},
  \label{eqn:a_sol}
\end{equation}
where $A$ and $B$ are constants of integration. 

Considering the special case of the inflating universe; during inflation $H\propto\dot{\phi}\sim\mathrm{const}$, so $z\propto a$. Thus, quantising the Mukhanov variable during inflation is equivalent to quantising a massless, minimally coupled scalar field on the same background spacetime. Note however, that in a more general scenario, the two spacetimes will not be the same.

\section{Minimising the renormalised stress-energy tensor}
Within the theory of quantum fields in curved spacetime, one is able to compute a {\em renormalised\/} stress-energy tensor $\bracket{0}{T_{\mu\nu}}{0}_\mathrm{ren}$. There are a variety of methods of doing this, but if carried out carefully they yield the same result.

\subsection{Hadamard point splitting}
We briefly recap the procedure for evaluating a renormalised stress-energy tensor via a Hadamard point splitting procedure.  
The Hadamard Green function is defined by:
\begin{equation}
  G^{(1)}(x,\prm{x}) =\frac{1}{2} \bracket{0}{\left\{ \phi(x),\phi(\prm{x}) \right\}}{0}.
  \label{eqn:green_function}
\end{equation}
The coincidence limit $\prm{x}\to x$ formally would yield the expectation $\bracket{0}{\phi^2}{0}$, but this is unfortunately divergent. The strategy therefore is to subtract off de-Witt Schwinger geometrical terms $G^{(1)}_\mathrm{DS}(x,\prm{x})$ which may be absorbed into a renormalisation of the ``bare'' constants $G_B$ and $\Lambda_B$. One then takes the coincidence limit to yield a non-divergent quantity. 

To form the stress-energy tensor from the Green function, one operates with a bi-scalar derivative function $D_{\mu\nu}(x,\prm{x})$:
\begin{align}
  \bracket{0}{T_{\mu\nu}(x)}{0}_\mathrm{ren} =& \lim\limits_{\prm{x}\to x} \mathcal{D}_{\mu\nu}(x,\prm{x}) \left[ G^{(1)}(x,\prm{x}) - G^{(1)}_\mathrm{DS}(x,\prm{x}) \right],
  \label{eqn:point_splitting}
  \\
  \mathcal{D}_{\mu\nu}(x,\prm{x})=& \frac{1}{2}\left( \nabla_{\mu}\nabla_{\prm{\nu}} + \nabla_{\prm{\mu}}\nabla_{\nu} \right) -\frac{1}{2} g_{\mu\nu} \nabla_\alpha \nabla^{\prm{\alpha}} \nonumber\\
  &+ g_{\mu\nu} \frac{1}{2}m^2.  \nonumber
\end{align}
The Hadamard Green function~(\ref{eqn:green_function}) using the mode expansion~(\ref{eqn:mode_expansion}) becomes:
\begin{align}
  G^{(1)}(x,\prm{x}) = 
  \int\frac{d^3 k\:}{{(2\pi)}^3a(\eta)a(\prm{\eta})}&
  \Big(
  \chi_\bk(\eta)
  \str{\chi_{\bk}}(\prm{\eta})e^{i\bk\cdot(\bx-\prm{\bx})}+\nonumber\\
  &
  \str{\chi_\bk}(\eta)
  \chi_{\bk}(\prm{\eta})e^{-i\bk\cdot(\bx-\prm{\bx})}
  \Big).
  \nonumber
\end{align}
Inserting this expression into~(\ref{eqn:point_splitting}) will yield an expression which depends on the specific choice of mode function $\chi_\bk$. We now regard this expression as a {\em functional\/} of the independent variables 
\begin{equation}
  \mathcal{X} = \{\chi_\bk,\str{\chi_\bk},\prm{\chi_\bk},\prm{\str{\chi_\bk}}\},
  \label{eqn:vars}
\end{equation}
and aim to minimise this with respect to the functions. Since $G^{(1)}_{\mathrm{DS}}$ does not depend on these variables, this term can be ignored for the purposes of extremisation. Further, the functional derivatives such as $\frac{\delta}{\delta \chi_\bk}$ commute with the limit expression, so in fact minimising the renormalised tensor with respect to the mode functions is equivalent to naively minimising the traditional stress-energy tensor~(\ref{eqn:SET}). Inserting the mode function~(\ref{eqn:mode_expansion}) into~(\ref{eqn:point_splitting}) and taking the coincidence limit, one finds:
\begin{align}
  \bracket{0}{T_{00}(x)}{0}_\mathrm{ren} = &\frac{1}{2}\int \frac{d^3 k}{{(2\pi)}^3 a^2} (\prm{\chi_\bk}-\frac{\prm{a}}{a}\chi_\bk)(\prm{\str{\chi_\bk}}-\frac{\prm{a}}{a}\str{\chi_\bk})
  \nonumber \\
  &+\left( k^2 + m^2a^2 \right)\chi_\bk {\chi_\bk}^\ast + \tilde{T},
\end{align}
where $\tilde{T}$ signifies the plethora of additional terms arising from the renormalisation process that have no dependence on the variables $\mathcal{X}$.
Minimising this with respect to $\mathcal{X}$ subject to the constraint~(\ref{eqn:normalisation_1}) yields the relations:
\begin{align}
  |\chi_\bk|^2 &= \frac{1}{2\sqrt{k^2+m^2a^2}}, \\
  \prm{\chi_\bk} &= \left( -i \sqrt{k^2+m^2a^2} + \frac{\prm{a}}{a} \right) \chi_\bk.
  \label{eqn:rn_condition_m}
\end{align}
%We note that since these conditions are derived from the unrenormalised stress-energy tensor, the result would be achieved for any renormalisation procedure.

\subsection{Application to the Mukhanov variable.}
As described earlier, in order to apply this formalism to the inflationary Mukhanov variable, one should take set $m=0$ and replace $a$ with $z$:
\begin{equation}        
  |\chi_\bk|^2 = \frac{1}{2k},\qquad
  \prm{\chi_\bk} = \left( -i k + \frac{\prm{z}}{z} \right) \chi_\bk.
  \label{eqn:rn_condition}
\end{equation}
This should now be compared with the more usual HD conditions~(\ref{eqn:hd_condition}). Deep inside the horizon (${k\gg -{\prm{z}}/{z}}$) these two initial conditions are equivalent, but yield very different answers for infra-red modes (small $k$).
The second of these equations may be re-written in a more illuminating form:
\begin{equation}
  \prm{\left( \frac{\chi_\bk}{z} \right)} = -i k \left( \frac{\chi_\bk}{z} \right),
\end{equation}
which suggests that the co-moving curvature $\mathcal{R}=v/z$ is set with a ``positive frequency mode'' independent from any spacetime variation.

  It is important to recognise setting these conditions at \(\eta_0\) is equivalent to forcing the universe into a vacuum state at that moment, but there is minimal theoretical guidance as to when this should be\footnote{Although Lasenby (private communication) has suggested using successive adiabatic approximations to pick the vacuum epoch as the moment when the field is most ``particle-like'' -- see Lasenby, A. Space Sci Rev (2009) 148: 329, section 4.1}. Indeed, there is little reason to imagine that the universe should be in a vacuum state at any given moment. However, these conditions could also be used to build a formalism of excited states.

It is also important to realise that this vacuum does not claim to be interpretable in terms of particles. It is merely the mode function that minimises the renormalised stress tensor. In the language of Hamiltonian diagonalisation, or adiabatic vacuums, it would be a superposition of ``particle states''.

Reference~\cite{vacuum_choice} provides a review (particularly in the appendix) of various choices of initial conditions analogous to~\eqref{eqn:hd_condition} and~\eqref{eqn:rn_condition}. It is interesting to note that the Danielsson vacuum~\cite{Danielsson,Danielsson2} bears a striking similarity to the renormalisation initial conditions~\eqref{eqn:rn_condition} we have derived, but is instead derived from phenomenological grounds by imposing initial conditions around a high energy cutoff.

\section{Renormalising the kinetically dominated universe}
\begin{figure}                   
  \centering
  \includegraphics[width=\columnwidth]{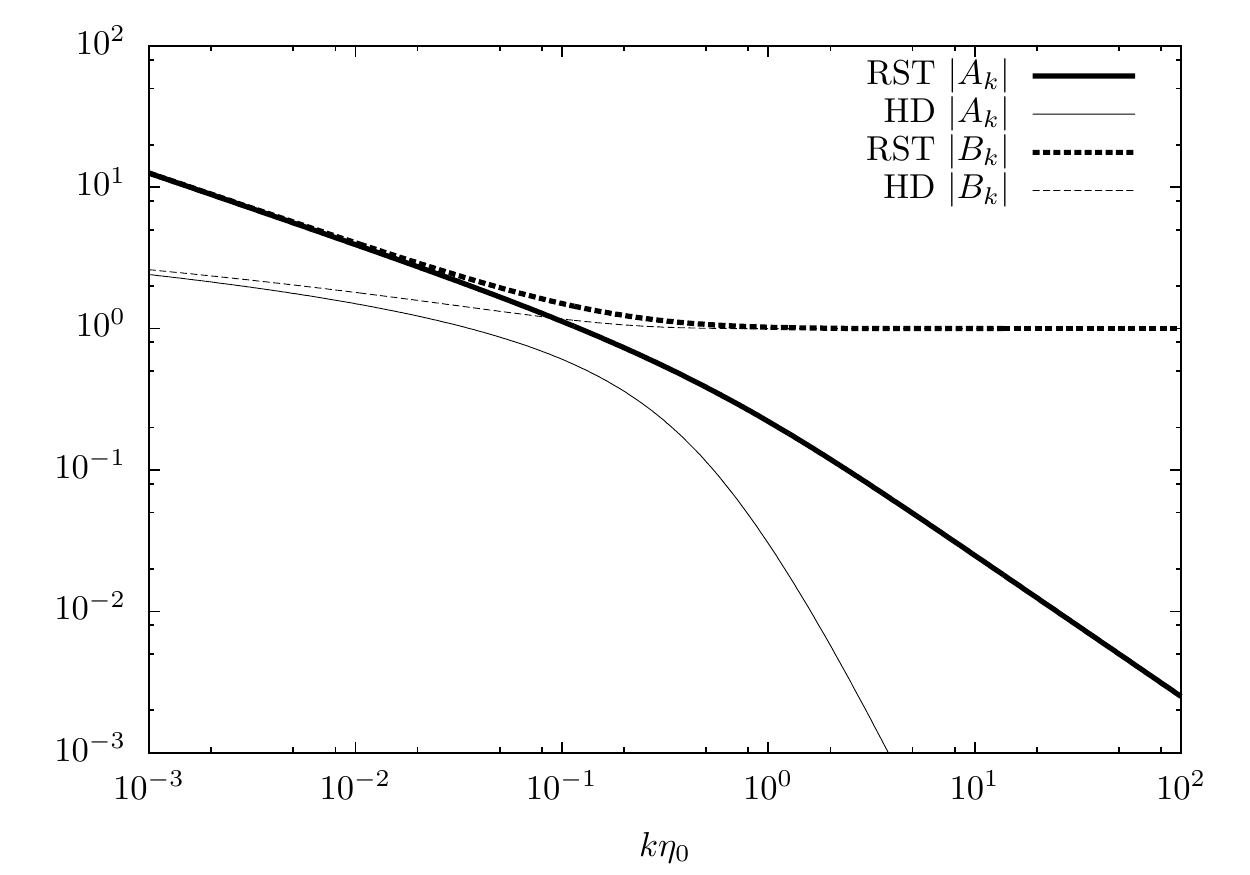}
  \caption{The modulus of the $A_k$ and $B_k$ coefficients in a kinetically dominated universe for the Hamiltonian diagonalising vacuum~(HD) and the vacuum from the renormalised stress tensor~(RST). Under these conditions, the universe will be in a vacuum state at conformal time $\eta_0$. Note that at large $k$, the mode functions tend to $A_k=0$, $B_k=1$.\label{fig:Ak}}
\end{figure}
We now consider these observations in the context of the kinetically dominated universe. 
It was recently observed~\cite{Handley+2014} that the classical solutions to the evolution equations~(\ref{eqn:klein_gordon})~\&~(\ref{eqn:Raychaudhuri}) emerge almost always from a kinetically dominated phase with $\dot{\phi}^2\gg V(\phi)$. 
In this regime, there is a significant period of cosmic time in which the theory of quantum fields in curved spacetime is valid. 
In this semi-classical pre-inflationary context, one finds that $\dot{\phi}\propto H$ and hence ${z\propto a}$. 
In the same manner as a de-Sitter universe, quantising the co-moving curvature perturbation is equivalent to quantising a massless scalar field on the same background spacetime. 
In this case though, the scale factor ${a \propto \eta^{1/2}}$, so the mode equations have the general solution:
\begin{align}
  \chi_\bk(\eta) &= \frac{1}{2}\sqrt{\pi\eta}\left( A_k H_{0}^{(1)}(k\eta) + B_k H_{0}^{(2)}(k\eta) \right), \nonumber \\
  1&=|B_k|^2 - |A_k|^2, \label{eqn:kd_modes} 
\end{align}
where without loss of generality we assume $A_k$ is real.
Applying HD conditions~(\ref{eqn:hd_condition}), or our new renormalised stress tensor conditions~(\ref{eqn:rn_condition}) yields different values for $A_k$ and $B_k$, as indicated in Figure~\ref{fig:Ak}. This difference is potentially observationally distinguishable, and will be analysed in a following paper.

\section{Conclusions}
We have presented a novel procedure for setting the initial conditions on the Mukhanov-Sazaki equation. We define the vacuum state via the instantaneous minimisation of the renormalised stress-energy tensor. This procedure is valid for any background cosmology, independent of the thorny issue of a particle-type concept. It reduces to the Bunch-Davies vacuum in an asymptotically static region. Further, it makes theoretical predictions that may be observationally testable.

\section*{Acknowledgements}
W.J.\ Handley would like to thank STFC for their support.

\bibliography{renormalisation}
\appendix

\end{document}